\documentclass[11pt]{article}
\usepackage{hyperref}
\pdfoutput=1
\usepackage[margin=1in]{geometry}

\begin{document}

\title{Drop Behavior in Uniform DC Electric Field}

\author{Paul F. Salipante and Petia M. Vlahovska \\
\\\vspace{6pt} Thayer School of Engineering \\ Dartmouth College, Hanover, NH 03755, USA}

\maketitle

\begin{abstract}
Drop deformation in uniform electric fields is a classic problem. The pioneering work of G.I.Taylor demonstrated that for weakly conducting media, the drop fluid undergoes a toroidal flow and the drop adopts a prolate or oblate spheroidal shape, the flow and shape being  axisymmetrically aligned with the applied field.  However, recent studies have revealed a nonaxisymmetric rotational mode for drops of lower conductivity than the surrounding medium, similar to the rotation of solid dielectric particles observed by Quincke in the 19th century. This fluid dynamics video, \href{http://ecommons.library.cornell.edu/bitstream/1813/14077/3/Drop_Behavior_Uniform_DC_FieldsMPG1.mpg}{Low Res} \href{http://ecommons.library.cornell.edu/bitstream/1813/14077/2/Drop_Behavior_Uniform_DC_FieldMPG2.mpg}{Hi Res}, demonstrates three behavioral modes.  I) toroidal recirculation inside the drop in weak fields II) nonaxisymmetric fluid rotation in strong fields and III) drop breakup  in strong fields.
\end{abstract}

\section{Physical mechanisms}
A charge-free drop with different physical properties than the surrounding fluid polarizes when placed in a uniform electric field. As a result, the electric stress changes discontinuously at the interface and deforms the drop.  In addition, electric field acting on induced free surface charges creates a tangential stress, not present in the case of perfect dielectrics, which drags the fluids into motion.

In weak electric fields, the suspending fluid undergoes axisymmetric extensional flow, which was explained by the leaky dielectric model introduced by G.I. Taylor.  He showed that the surface charge distribution and direction of surface fluid motion depend on the charging response of the fluids. If the charging time of the drop fluid is shorter than the suspending liquid, the interface charge distribution is dominated by charges brought from the interior fluid and the induced dipole moment is aligned with the electric field.  In this case, charges at the poles are attracted by the electrodes, pulling the drop into a prolate shape.  If the charging response of the suspending fluid is faster than the interior fluid, the charging of the interface is dominated by the exterior medium and the drop dipole is reversed; it is directed oppositely to the applied electric field.  In this charge configuration, the drop can become either prolate or oblate.

In strong electric fields, the flow pattern changes to rotation and drop deformation is no longer axisymmetric.  The electro-rotation behavior of drops is analogous to the spontaneous spinning of a rigid sphere in a uniform electrostatic field, reported by Quincke in 1896.  For rotation to occur, the suspending fluid and sphere must be slightly conducting dielectrics and have the induced dipole moment of the sphere oriented in opposite direction of the applied field.  This configuration becomes unstable above a critical strength of the electric field.  A perturbation in the dipole alignment induces physical rotation of the sphere.  The charge distribution rotates with the sphere, however, the exterior fluid simultaneously recharges the interface.  The balance between charge convection by rotation and interface charging results in an oblique dipole orientation.  The rate of rotation and threshold electric field are determined from angular momentum conservation.

\section{Experiments}

Silicone oil drops suspended in castor oil are subjected to a uniform DC electric field.   A fluid chamber with flat plate brass electrodes is used to observe drop behavior.  The drop size, viscosity ratio, and electric field strength  are varied to produce different behavior.  The electrical conductivity, dielectric constant, and viscosity are determined for each fluid in order to evaluate the system parameters for conductivity ratio R=$\sigma_{in}/\sigma_{out}$, permittivity ratio S=$\epsilon_{out}/\epsilon_{in}$, and viscosity ratio $\lambda$=$\mu_{in}/\mu_{out}$.
The electric conductivities of the castor oil and silicone drops are $\sigma_{out}$=4.5x10$^{-11}$S/m $\sigma_{in}$=1.2x10$^{-12}$S/m and the relative permittivities are $\epsilon_{out}$=5.3 and $\epsilon_{in}$=3.0, respectively.  The viscosity of the continuous fluid is $\mu_{out}$=0.69 Pa-s and the viscosity of the drop phase is varied from $\mu_{in}$=0.05 Pa-s to $\mu_{in}$=9.74 Pa-s.  The videos are shown in real time speed (15 frames/second).

\subsection{Toroidal Flow}

A 3.4mm diameter silicone oil drop with $\lambda$=0.5 (one half the viscosity of the suspending castor oil medium)  is subjected to a 3.6 kV/cm uniform DC electric field.  To observe the flow field, the drop is seeded with 10$\mu m$ diameter aluminum particles.  
\subsection{Transition to Oblique Orientation}

A 2.25mm diameter silicone oil drop with $\lambda$=0.3 is subjected to a uniform DC electric field with increasing strength.  The transition to oblique orientation with respect to the electric field is observed above a critical field strength of 6.5 kV/cm, approximately 2.3 times the critical field predicted by the Quincke expression.

\subsection{Breakup}

A 5.7mm diameter silicone oil drop with $\lambda$=1.4 is subjected to a 5.1 kV/cm uniform DC electric field.

 \end{document}